# Blockchain Function Virtualization: A New Approach for Mobile Networks Beyond 5G

Shiva Kazemi Taskou, Mehdi Rasti, and Pedro H. J. Nardelli

*Abstract*—Many of the key enabling technologies of the fifth-generation (5G), such as network slicing, spectrum sharing, and federated learning, rely on a centralized authority. This may lead to pitfalls in terms of security or single point of failure. Distributed ledger technology, specifically blockchain, is currently employed by different applications related to the Internet of Things (IoT) and 5G to address the drawbacks of centralized systems. For this reason, mobile blockchain networks (MBNs) have recently attracted a great deal of attention. To add a transaction to the blockchain in MBNs, mobile or IoT users must perform various tasks like encryption, decryption, and mining. These tasks require energy and processing power, which impose limitations on mobile and IoT users' performance because they are usually battery powered and have a low processing power. One possible solution is to perform the tasks virtually on commodity servers provided by mobile edge computing (MEC) or cloud computing. To do so, all tasks needed to add a transaction to the blockchain can be treated as virtual blockchain functions that can be executed on commodity servers. We introduce a blockchain virtualization framework called blockchain function virtualization (BFV), through which all blockchain functions can be performed virtually by MEC or cloud computing. Furthermore, we describe applications of the BFV framework and resource allocation challenges brought by the BFV framework in mobile networks. In addition, to illustrate the advantages of BFV, we define an optimization problem to simultaneously minimize the energy consumption cost and maximize miners' rewards. Finally, simulation results show the performance of the proposed framework in terms of total energy consumption, transaction confirmation rate, and miners' average profit.

## I. INTRODUCTION

Blockchain is a distributed ledger technology providing a decentralized peer-to-peer network in which data are added to the blocks in the form of transactions after verification, and blocks are added to the blockchain after verification and consensus [1]. Once a block is added to the blockchain, transactions cannot be modified or deleted (immutability), and the history of all transactions is placed in a unique blockchain [2]. Without relying on a centralized authority, blockchain can alleviate the problems of centralized systems, such as a single point of failure, security, vulnerability to an outage, and a high probability of attacks.

On the other hand, most technologies of 5G and also beyond 5G still rely on a centralized authority to ensure data integrity, trust, and security. Such a centralized system architecture is not the only option; blockchain may provide a feasible alternative to build a secure and trustworthy network with a distributed architecture. In this context, mobile blockchain networks (MBNs) have attracted much attention from industry and academia, MBNs still face several technological challenges to be deployed on a large scale. For example, an MBN requires a high processing power and storage for recording transactions, and has a high energy consumption.

This article addresses the challenges of the MBN by developing a blockchain function virtualization (BFV) framework in which blockchain functions, such as mining, encryption, and decryption, are treated as virtual blockchain functions that can be performed on commodity servers by mobile edge computing (MEC) or cloud computing. To illustrate the advantages of the proposed approach, we propose an optimization problem to minimize the energy cost and maximize miners' rewards, which is the basis of a series of numerical results presented to demonstrate the high performance of BFV.

This article has the following main contributions:

- In the previous related works, only the mining process is offloaded to commodity servers in the cloud or MEC, and the lack of sufficient memory and the processing power limitation on performing other blockchain functions are not addressed. In contrast, in the BFV framework, not only the mining process but also the other blockchain functions (e.g., encryption and decryption) can be virtually performed on commodity servers. By doing so, BFV makes battery–powered mobile and IoT users participate in blockchain networks. In this article, we also demonstrate applications of BFV and pin–point resource allocation challenges caused by BFV in mobile networks.
- To address the energy consumption challenge within such a BFV framework, we define an optimization problem that aims to simultaneously minimize the energy consumption cost and maximize miners' rewards. To address this problem, we employ the majorization–minimization approximation method [3].
- Through simulation results, we confirm that by the virtually performing blockchain functions, the proposed BFV framework achieves lower energy consumption, a higher transaction confirmation rate, and a higher miners' reward compared with the previous works that only offloads the mining process.

## II. OVERVIEW OF MOBILE BLOCKCHAIN NETWORKS

The transactions of the MBNs (which are generated by mobile and IoT users) are placed in blocks after verification.

S. Kazemi Taskou and M. Rasti are with Department of Computer Engineering, Amirkabir University of Technology, Tehran, Iran. M. Rasti is also a visiting researcher at Lappeenranta-Lahti University of Technology, Lappeenranta, Finland. (email: {shiva.kt,rasti}@aut.ac.ir).

Pedro H. J. Nardelli is with Lappeenranta-Lahti University of Technology and also with University of Oulu, Finland. (e-mail: Pedro.Nardelli@lut.fi)

Miners generate these blocks, and a miner that completes the mining process faster than the other miners will broadcast the block to all other nodes for verification. This block is added to the blockchain after verification [1]. In the following, we explain the steps to create a transaction and add thatto the blockchain in more detail.

## A. Transaction Generation and Broadcasting

In the MBN, to generate a transaction, each user creates a public and a private key by the RSA algorithm. The user then places the required information in the transaction. For example, in federated learning, if the user wants to share their local learning model with others, they should place their local learning model in the transaction. Each transaction has a unique ID for identification. To obtain this ID, the hash of all the transaction information must be taken. Then, to authenticate and verify the transaction, the transaction generator must sign the transaction. To sign the transaction, the hash value from all the transaction information is first taken, and then the hash value is encrypted with the private key of the transaction generator. After performing these steps, this transaction will be broadcast to the other users for verification and authentication.

For broadcasting the transactions and blocks in blockchain networks, a gossip protocol is employed [1]. In the gossip protocol, any user that receives transactions or blocks sends them to all their neighboring users. In wireless networks, all users should broadcast transactions and blocks through wireless communication. Because of the size of transactions and blocks in the MBN, users may require a lot of energy to broadcast transactions and blocks, and the battery-powered mobile and IoT users may not have enough energy for broadcasting.

In summary, as shown in Fig. 1, the user who wants to perform a transaction should implement two functions, i.e., transaction generation and transaction broadcasting. The transaction generation function includes the subfunctions as illustrated in Fig. 1.

## B. Block Generation, Mining, and Block Broadcasting

After transaction broadcasting, each user who receives this transaction authenticates and verifies it, and then stores it in their memory. To authenticate a transaction, the transaction receiver user decrypts the digital signature of the transaction using the sender's public key. And to verify the transaction, the user calculates the hash of the transaction. The user then places several transactions stored in their memory in a block.

The miners that tend to generate blocks perform the mining process to add a block to the blockchain, through which each miner selects a nonce that should be placed at the block header. When a correct nonce is found, the generated block will be broadcast to all other users for verification [1].

As shown in Fig. 1, any miner who is interested in block generation must perform blockchain functions, including authentication, verification, block generation, and mining functions. Then, the generated block is broadcast to all other network nodes. Because the mining function requires a lot of processing resources, the mining process can be performed in parallel on multiple servers.

## C. Adding a Block to a Blockchain

Users that receive the generated block verify the block and then add it to the blockchain stored in their memory. To add a block to a blockchain, the hash of the last block in the blockchain is inserted in the header of the new generated block. Fig. 1 illustrates the block generation and broadcasting functions.

## III. APPLICATIONS AND CHALLENGES OF THE MBN IN 5G

This section highlights some key applications and challenges of the MBN in 5G (and beyond 5G) wireless networks. For more information and other applications, interested readers are referred to [4].

**Network slicing:** As one of the key enabling technologies for 5G and beyond, network slicing enables operators to share a common infrastructure with multiple virtual operators and tenants to provide services to the end users. A broker performs the network slicing based on a centralized architecture and may suffer from issues related to centralized systems. The MBN can solve the problems of a centralized broker by implementing its role in a distributed manner [5]. In blockchain-enabled network slicing, each user announces their required resources and quality of service (QoS) through transactions. Furthermore, users can be made aware of unutilized resources and the amount of resources allocated to them by referring to the blockchain.

**Roaming:** When a user connects to the visited operator from the home operator in roaming, they should pay a roaming charge to the visited operator. A malicious user may not pay the roaming charge and bypass the system. The MBN can anticipate and prevent these fraudulent activities [5] by storing the history of the user's connectivity and such a payment in the blockchain.

**Frequency spectrum sharing:** Currently, the frequency spectrum sharing among operators is accomplished through a centralized regulatory authority [5]. This spectrum sharing method has several drawbacks, including a single point of failure, trust, and security. The MBN can address those issues by recording the information of frequency spectrum allocation to all operators and their payments in the blockchain.

**IoT:** Typically, the IoT consists of millions of simple devices, such as sensors, aggregators, and decision triggers, that may interact with each other. These devices are usually provided by different service operators and vendors, requiring mutual trust. Furthermore, the integrity of data exchanged between devices should be ensured. Currently, for user authentication and access control, IoT networks need a trusted centralized authority that suffers from a single point of failure, trust issue, and scalability problems. Because of the huge number of IoT devices, this centralized authority causes a bottleneck leading to congestion and packet drop. The MBN can overcome these drawbacks [6]. In some use cases of the IoT, such as health care, sensitive data of users are stored in a centralized database [6]. This centralized database can suffer from a single point of failure, trust issues, and security attacks. The MBN can address many of these challenges [6]. Because in the MBN, IoT users' data are stored in a blockchain obtained by a distributed consensus mechanism, data integrity is assured. Relying on public keys, the MBN can perform authentication and access control for IoT users.

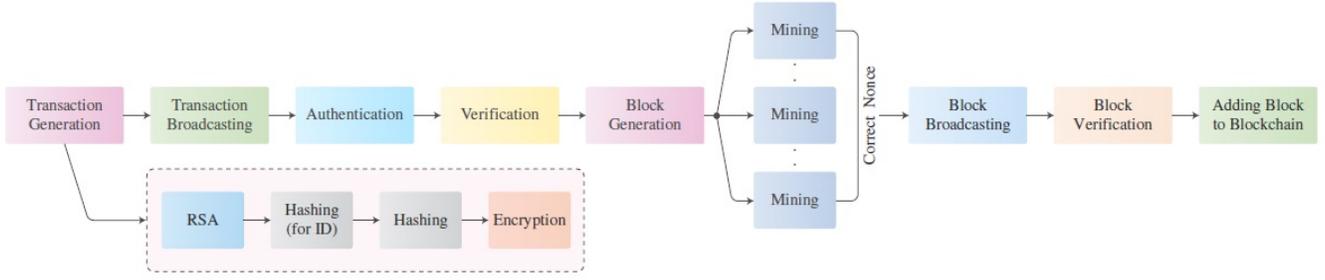

Fig. 1: Blockchain functions required to add a block to a blockchain

**Vehicular networks:** Communication between vehicles plays an essential role in intelligent transportation systems. The most important requirement of vehicle networks is trust. The MBN can build trust between distributed vehicles without a centralized authority [7]. Moreover, the MBN can overcome the challenge of vehicular networks, such as vulnerability to various attacks.

**Federated learning:** Federated learning is a distributed learning method on wireless networks in which users learn a local model by using their local data. The users then send their local models to a centralized entity. The centralized entity learns the global model by using the local models received from the users and sends the global model to the users. In federated learning, in addition to the problem of a single point of failure, malicious users may send fake local models to the centralized entity [8]. The MBN allows users to share their local model by transactions. Further, each user can access the other users' local model by referring to the blockchain.

In the MBN, mobile and IoT users may also generate transactions depending on the application. To verify the transactions, mobile and IoT users should generate a new block by performing the mining process. The mining process requires a high processing power and consumes a significant amount of energy, and thus, mobile and IoT users cannot perform the mining process directly because of their limited processing power and battery-powered energy. This drawback poses a major challenge in applying blockchain to mobile networks and IoT. In addition to the processing intensiveness of the mining process, the MBN has other challenges. First, mobile and IoT users may not have enough energy and processing power to run encryption, decryption, and hash algorithms [9]. In addition, they may not have sufficient memory to store verified transactions and the blockchain [1], making memory another key limitation.

Moreover, scalability and different QoS provisioning pose other fundamental challenges. Scalability is defined as the number of confirmed transactions per second, i.e., the transaction confirmation rate. On the other hand, different MBNs have different QoS; e.g., the blockchain for the IoT should be low cost and guarantee privacy, while an MBN for federated learning should assure low latency. Hence, mobile networks and IoT need to be designed to support several blockchains with diverse QoS [12].

To realize an MBN, the scalability of the blockchain should be improved, and multiple blockchains having different QoS requirements should coexist on a common infrastructure. To do so, the blockchain network should be provided with both processing power and memory required for performing encryption, decryption, hash algorithms, and the mining process, and storing transactions and the blockchain.

## A. STATE OF THE ART

To realize an MBN, frameworks for integrating cloud computing and MBNs were proposed in [10]– [11] to perform the mining process on general purpose servers. Because of the higher latency of cloud computing, recently, MEC has been used for implementing MBNs. For instance, authors in [2] proposed a framework for integrating MEC and an MBN. All the frameworks proposed in [2] and [10]– [11] only deal with users' processing power limitation to perform the mining process by offloading it to the cloud or MEC servers. However, they do not address the lack of sufficient memory and processing power limitation on performing other tasks, including encryption, decryption, and hash algorithms for transaction generation and verification of blocks and transactions.

Furthermore, in [12], a virtualization framework was proposed to facilitate the implementation of various blockchains on a physical infrastructure and alleviate scalability challenges. In this framework, the tasks of blockchain nodes are virtually implemented in the cloud. In addition, a microservice-based framework has been proposed [13] through which blockchain nodes can perform tasks related to different blockchains as one or more microservices in the cloud.

In this article, we take one step further and propose a framework in which the required processing power for performing mining, encryption, decryption, and hash algorithms and memory to store transactions and the blockchain are provided by commodity servers in MEC or the cloud. Inspired by virtualization in 5G and beyond provided to support heterogeneous services on a common infrastructure, virtualization can also be used to tackle the aforementioned blockchain challenges. In a blockchain, a chain of functions must be performed sequentially to add a block to the blockchain. Virtual performing of these functions has some benefits for the blockchain, including provision of the required processing power and memory, improved scalability, flexibility, reduced energy consumption, faster implementation of new blockchains, and deployment of diverse blockchains on a common infrastructure.

In the next section, we propose a blockchain virtualization framework named BFV, in which all tasks of the MBN are performed as virtual blockchain functions on commodity servers (in contrast to [2], in which only the mining task is offloaded to MEC servers).

## IV. PROPOSED BFV FRAMEWORK AND ITS APPLICATIONS

Inspired by network function virtualization, in which network functions are decoupled from hardware and virtually performed on commodity servers [14], all blockchain functions can be performed on commodity servers in MEC and the cloud. Through the proposed framework, BFV, as shown in Fig. 2, each of the blockchain functions is performed on commodity servers, and the output of each function is sent through physical links to the server that executes the next function. In BFV, all functions illustrated in Fig. 1 are performed on commodity servers as virtual blockchain functions.

*Initialization setup in BFV*: In BFV, when users enter the network, each of them is associated with an available base station. Then, if they want to send or receive transactions, they generate a blockchain account. To do so, they generate a pair of public and private keys via the RSA algorithm. The users can generate these keys locally or offload the required processing to a general purpose server in the cloud or MEC through the associated base station. Furthermore, if users desire to store a copy of the blockchain, upon entering the network, they request the last copy of the blockchain from the neighbors. At the initialization of a blockchain, the blockchain has a genesis block, which is the first block in each blockchain, and it does not include any transaction. The users can store a copy of the blockchain either locally in the storage of their device or remotely in the available storage in the cloud or MEC.

As shown in Fig. 2, to generate a transaction in the BFV framework, a user transmits the information that should be inserted in a transaction to the commodity servers via base stations. The transaction generation function is then performed on an appropriate server allocating the required processing resources and storage. The generated transaction is then sent to a server that performs transaction broadcasting via physical links. This server sends the generated transaction to all users in the MBN.

Users who receive the transactions and tend to perform the mining process send their block generation requests to the commodity servers. As indicated in Fig. 1, the virtual functions of authentication, verification, and block generation are then performed sequentially. After this step, a new block is generated, which enters the selected servers for the mining process through physical links. Several servers may be selected to perform the mining process in parallel. After finding the correct nonce for the new block, the block broadcasting function is executed, through which the generated block is broadcast to all users of the MBN.

Each user that receives this block first executes the block verification function and then adds the block to their blockchain. Note that servers that implement the transaction generation and adding the block to the blockchain functions should have enough storage to store verified transactions and the blockchain.

### A. APPLICATIONS OF BFV

BFV facilitates the applications of MBN and realizes blockchain as a service (BaaS). Because of the complexity of blockchain deployment, many developers are unable to deploy their specific blockchains. For this purpose, BaaS provides the required infrastructure and resources for developers to host, run, and manage their blockchains [15]. Through BFV, mobile operators can offer BaaS to developers and thus boost their revenue.

## V. OPEN ISSUES AND RESEARCH DIRECTIONS

The proposed BFV opens up new research challenges for resource allocation, changing more established optimization problems in terms of their objective function, constraints, and decision variables. Specifically, these differences based on their effects on optimization problems are as follows:

- *Objective function:* Maximization of transaction confirmation rate (defined as the number of confirmed transactions per second) is a particular objective function in BFV. Furthermore, in BFV, to obtain rewards from a blockchain, miners tend to maximize their winning probability.
- *Constraints:* The orphaning and forking probability should be taken into account in the optimization problem. Forking means that several miners win the mining process simultaneously, resulting in several blockchains in the network. Orphaning means that if a block is not accomplished within the set deadline, the generated block becomes an orphan block. In BFV, the probability of becoming an orphan block depends on wireless communication characteristics and implementation of the blockchain functions on commodity servers.

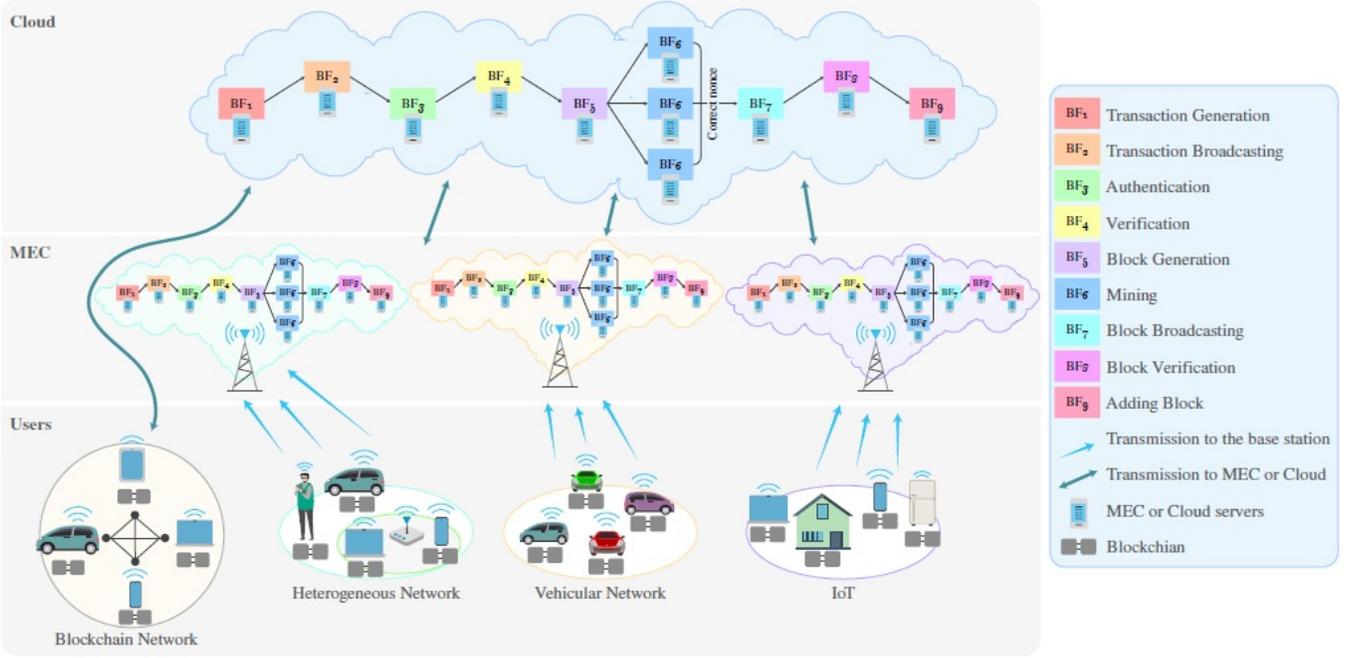

Fig. 2: Framework of blockchain function virtualization (BFV can be run on MEC or cloud servers or hierarchically on MEC and cloud servers)

- *Decision variables:* In BFV, the block size affects the orphaning probability, reward, and transaction confirmation rate. A larger block size results in a higher orphaning probability, reward, and transaction confirmation rate. Thus, the block size can be considered a decision variable. On the other hand, miners can perform the mining process in two modes: (1) carrying out the mining process individually (solo mining) or (2) cooperating with other miners (pool mining). Hence, mining mode selection can be considered a binary decision variable.

To maximize the transaction confirmation rate by BFV, users must be able to send more transaction generation requests per second to generate and broadcast more transactions. To this end, the high data rate requirements should be guaranteed. These requirements are met through the enhanced mobile broadband (eMBB) service on 5G and beyond. Furthermore, in some applications such as IoT, BFV must be able to generate and broadcast transactions for a massive number of users, and thus, the massive machine type communication (mMTC) requirement should be met. Furthermore, low latency and high reliability requirements related to reliable low latency communication (RLLC) services in mobile networks must be guaranteed to reduce the orphaning probability. Therefore, for instance, an MBN designed for IoT applications by applying the BFV framework, must support all three types of eMBB, mMTC, and RLLC services simultaneously. Assuring different service requirements is an open problem in (beyond) 5G networks.

### A. Challenges of Radio Resource Allocation in the RAN

In BFV, users and miners transmit their transaction and block generation requests to commodity servers through base stations. The communication between the users and the base stations is provided by wireless communication, and therefore, resource allocation in the radio access network (RAN) is an important challenge. Resource allocation in the RAN including user association, power control, and frequency spectrum allocation should be performed to satisfy users' and miners' requirements.

### B. Challenges of Blockchain Function Placement and Routing

The commodity servers in MEC or the cloud differ in terms of processing power and storage. These servers communicate with each other through physical links. For BFV to be cost-effective for mobile network operators and reduce energy consumption, allocation of the commodity servers (called blockchain function placement) and the allocation of physical link bandwidth (called routing) must be efficiently performed.

Moreover, in BFV, as shown in Fig. 2, the blockchain functions can be performed on commodity servers in MEC or the cloud. They can also be run hierarchically on MEC and cloud servers. There is a conflict between processing power and delay in MEC and the cloud. Specifically, owing to the higher processing power of the cloud in comparison with MEC, the processing delay in the cloud is less than that of MEC. However, the RAN delay for cloud computing is more than for MEC because the distance between the users and the cloud is more than the distance between the users and MEC. Therefore, the RAN delay in the cloud is more than in MEC. With respect to the available processing power in MEC and the cloud and the deadline for adding a block to the blockchain, the decision about whether a blockchain function performs in MEC or the cloud is an interesting topic left for future work.

## C. Challenges of Resource Allocation in Blockchain

There are different performance criteria in BFV, such as energy consumption, transaction confirmation rate, and orphaning probability. There is a trade-off between the probability of orphaning and transaction confirmation rate in BFV so that a larger size block leads to a greater transaction confirmation rate and a higher probability of orphaning. Allocation of RAN resources, processing resources, and bandwidth of physical links should be performed so that a block is completed on time to prevent the orphaning. Resource allocation methods should also strike a balance between the probability of orphaning and the transaction confirmation rate. Moreover, latency and energy consumption in BFV are affected by radio resource allocation in the RAN and blockchain function placement and routing in MEC or the cloud; therefore, the joint allocation of such resources is an important problem.

## D. Resource Allocation Challenges of Integration of BFV and Other Technologies

When BFV is integrated with other technologies in 6G, such as the IoT, vehicular networks, and federated learning, new resource allocation challenges arise. One of the most important challenges is that resource allocation must be performed so that in addition to meeting the QoS requirements of users deploying these technologies, the blockchain network requirements must also be met. Moreover, BFV can offer BaaS. For this purpose, resources should be allocated to each blockchain network so that the QoS requirements of each blockchain network are satisfied, and a change in one blockchain does not affect the performance of other blockchain networks.

## VI. ENERGY-EFFICIENT RESOURCE ALLOCATION FOR BFV

To exemplify an application of the proposed framework, we consider an MBN consisting of a set of $\mathscr{U} = \{1, 2, \cdots, U\}$ users. Any user who wants to generate a transaction or participate in the mining process transmits their request to the base station. We model the MEC or the cloud as a directed graph $\mathscr{G} = (\mathscr{N}, \mathscr{L})$, where $\mathscr{N}$ is a set of servers, and $\mathscr{L}$ is a set of directed links. In MEC or the cloud, server processing resources are allocated to execute each function of the requests. The result of each function is sent via physical links to the server responsible for executing the next function. In this network, users and miners consume energy in the RAN to transmit their requests to the base station, and the commodity servers consume energy to perform blockchain functions. Furthermore, maximizing the reward is an important objective for miners. Thus, we aim to simultaneously minimize the energy cost (denoted by $E_{\text{total}}$) and maximize the miners' rewards (denoted by $R_{\text{mining}}$). Let $x_n^{i,j}$ denote the server allocation binary variable. The blockchain function placement problem for adding each block to the blockchain is formally stated as

$$\min_{\{x_n^{i,j}\}} E_{\text{total}} - R_{\text{mining}}$$

s.t.

$$\text{C1}: \quad T^{\text{RAN}} + T^{\text{MEC}} \leq T^{\text{th}}, \quad (1)$$

$$\text{C2}: \quad \sum_{i \in \mathscr{U}} \sum_{j \in \mathscr{S}_i} x_n^{i,j} C_{i,j} \leq C_n^{\max}, \quad \forall n \in \mathscr{N},$$

$$\text{C3}: \quad x_n^{i,j} \in \{0, 1\}, \quad \forall i \in \mathscr{U}, \quad \forall j \in \mathscr{S}_i, \quad \forall n \in \mathscr{N},$$

where $\mathscr{S}_i$ is a set of the user $i$'s requested functions. $T^{\text{RAN}} + T^{\text{MEC}}$ is the time taken for generation of a block to broadcast it. Particularly, $T^{\text{RAN}}$ represents the delay for transmitting the miners' requests to the base stations and responses from the base stations to the users calculated as RAN transmission time [3]. $T^{\text{MEC}}$ denotes the summation of processing delay on servers for performing blockchain functions. $T^{\text{MEC}}$ is obtained by $T^{\text{MEC}} = \sum_{n \in \mathscr{N}} \sum_{j \in \mathscr{S}_i} x_n^{i,j} C_{i,j} / C_n^{\max}$, where $C_n^{\max}$ denotes the processing capacity of the server $n$ and $C_{i,j}$ denotes the required CPU cycles for performing the $j$th function of $\mathscr{S}_i$. Moreover, $T^{\text{th}}$ represents the block interval, meaning that blocks should be generated and broadcast at each interval of $T^{\text{th}}$ seconds, and otherwise, the block will be orphaned.

The total energy consumption for adding a block to the blockchain is obtained by

$$E_{\text{total}} = E_{\text{RAN}} + E_{\text{MEC}}, \quad (2)$$

where $E_{\text{RAN}}$ is the energy consumption for transmission requests to the base station. $E_{\text{MEC}}$ is energy consumption for the implementation of all blockchain functions on the commodity servers. Similar to [3], $E_{\text{MEC}}$ is given by $E_{\text{MEC}} = \sum_{n \in \mathscr{N}} \widetilde{p}_n (\sum_{i \in \mathscr{U}} \sum_{j \in \mathscr{S}_i} x_n^{i,j} C_{i,j} / C_n^{\max})$, where $\widetilde{p}_n$ represents power consumption of the server $n$ for processing per each second.

Successful mining of a block depends on the probability of finishing the mining faster than the others and the orphaning probability. The probability of finishing the mining faster than the others denoted by $p_{\text{mining}}$ is obtained from the ratio of the miner user $i$'s demand to the total demands of all miner users [3]. Further, the probability of a block orphaning is given by $p_{\text{orphan}} = 1 - e^{-\lambda z N_{\text{Trans}}}$, where $\lambda = \frac{1}{T^{\text{th}}}$, $z$ is a given network latency parameter, and $N_{\text{Trans}}$ is the number of transactions in the mined block. Thus, the corresponding reward to the miner $i$ is obtained by $R_{\text{mining}} = p_{\text{mining}}(R_{\text{const}} + N_{\text{Trans}} R_{\text{Trans}})(1 - p_{\text{orphan}})$, in which $R_{\text{const}}$ is a constant reward and $R_{\text{Trans}}$ is a reward of each transaction.

The problem (1) is an integer linear problem, and its optimal solution cannot be obtained in polynomial time. To tackle this difficulty, we convert the binary variable $x_n^{i,j}$ into a continuous one and add a penalty function as

$\lambda \sum_{i \in \mathscr{U}} \sum_{j \in \mathscr{S}_i} \sum_{n \in \mathscr{N}} \left( x_n^{i,j} - (x_n^{i,j})^2 \right)$ to the objective function. This penalty function makes the problem (1) nonconvex. Therefore, to convexify the problem, we use the majorization–minimization approximation method [3] to approximate $(x_n^{i,j})^2$. This approach converts problem (1) into a linear problem, which can be optimally solved by off-the-shelf optimization packages.

## A. SIMULATION RESULTS

To evaluate the performance of the proposed BFV framework, we compare it with the framework presented in [2], in which only the mining process is offloaded to MEC. To do so, we consider the total energy consumption for adding each block to the blockchain, the transaction confirmation rate, and miners' average reward as the performance measurement criteria. Similar to Bitcoin, we assume that for transaction generation, the RSA, SHA-256, and ECDSA algorithms; for block authentication, the SHA-256 and ECDSA algorithms; and for adding a block to the blockchain, the SHA-256 algorithm are employed.

The required processing power for the aforementioned algorithms is given according to the information presented in https://www.cryptopp.com/benchmarks.html. The simulation parameters are given in Table I.

TABLE I: Simulation Parameters

| Parameter | Value |
|---|---|
| Number of MEC servers | 50 |
| Required CPU cycles for SHA-256 | 15.8 CPU cycles/s for each byte |
| Required CPU cycles for RSA | $36 \times 10^6$ CPU cycles/s |
| Required CPU cycles for ECDSA | $5.27 \times 10^6$ CPU cycles/s |
| Energy consumption of gossip broadcasting | 12.5 Jule |
| Required CPU cycles for block authentication | 15.61 CPU cycles/s for each byte |
| Required CPU cycles to make a Merkle tree | 15× required CPU cycles for SHA-256 |
| Required CPU cycles for the mining process | $0.25 * 10^9$ CPU cycles/s |
| power consumption for processing each CPU cycle ($\widetilde{p}_n$) | 125 W |
| Power consumption of users for transmission per second | 0.2 W |
| Processing capacity of IoT sensors | 0.01 GHz |
| Processing capacity of mobile users | 0.1 GHz |
| Processing capacity of MEC servers | 5 GHz |
| Block interval ($T^{\text{th}}$) | 1 s |

For comparison with [2], we assume that the total energy consumption to add a block to the blockchain in [2] consists of (1) the energy consumption of users for performing all blockchain functions shown in Fig. 2 except the mining process; (2) the energy consumption for transmission of the mining task to the base station; and (3) the energy consumption of processing the mining task in MEC servers.

In Figs. 3a and 3b, the total energy consumption for adding a block to the blockchain in BFV and the framework proposed in [2] are compared in terms of the servers' processing capacity and the block size. Moreover, Figs. 3c and 3d illustrate the transaction confirmation rate in BFV and the framework proposed in [2] versus the number of miners and the block size. Furthermore, Figs. 3e and 3f depict the miners' average reward in BFV and the framework proposed in [2] versus the number of miners and a different block interval.

To generate Figs. 3a, 3c, 3e, and 3f, we assume that there are 5,000 transactions in each block and the size of each transaction is 200 bytes. Further, to generate Figs. 3a, 3b, 3d, and 3f, the number of miners in the network is set to 50, where one-third of them are IoT sensors and the rest are mobile users.

As can be seen in Fig. 3a, as the servers' processing capacity increases because more processing capacity is available to perform blockchain functions, the total energy consumption decreases. In addition, it is observed that because in BFV, users and miners perform all the required functions virtually in MEC, the energy consumption is reduced compared with the proposed framework in [2], in which only the mining process is offloaded to the MEC servers. Moreover, it can be observed from Fig. 3b that the total energy consumption is increased when the block size increases. Because in MBN, energy is consumed for each transaction added to the blockchain, as the number of transactions on each block increases, the total energy consumption also increases.

Fig. 3c shows that when the number of miners increases, the transaction confirmation rate decreases. Because the processing capacity of the servers in MEC is limited when more users share these limited resources, the time to add a transaction to the blockchain will also increase. Hence, the transaction confirmation rate decreases. Moreover, as shown in Fig. 3d, the number of confirmed transactions increases as the block size increases. The reason is that when the number of transactions per each block is increased, more transactions are added to the blockchain as a result of each successful mining process.

In Fig. 3e, it can be observed that with an increasing number of miners, the average miner reward decreases, because the processing resources are shared between more miners, and the miners' rewards depend on the demand of all miners. In addition, Fig. 3f illustrates that when the block interval increases as the orphaning probability decreases, the average miners' reward increases.

## VII. CONCLUSION

Because of the expected increase in blockchain applications in 5G and beyond 5G networks, it is necessary to address the related implementation challenges, such as required energy, processing power, memory, and scalability. To alleviate these challenges, this article provided a BFV framework for performing blockchain functions as virtual functions on general purpose servers. As mentioned above, BFV can facilitate various blockchain applications in mobile networks, e.g., IoT, vehicle networks, and federated learning. However, to enjoy the benefits of BFV, its challenges, such as resource allocation, must be overcome. In this article, we defined a blockchain function placement problem to minimize energy cost and maximize miners' reward. The simulation results confirm that the proposed BFV framework outperforms a reference framework in which only the mining process is offloaded to MEC servers in terms of energy consumption, transaction confirmation rate, and miners' average reward.

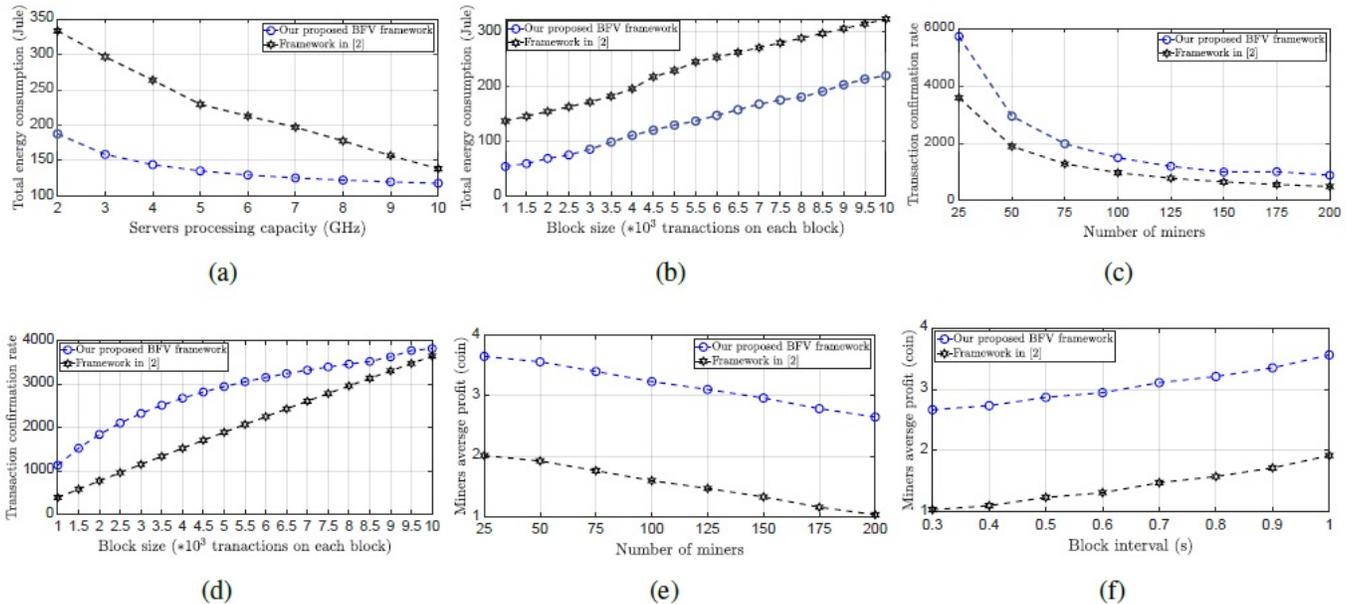

Fig. 3: Comparison of BFV and the framework proposed in [2] in terms of (a) total energy consumption vs. processing power of the servers; (b) total energy consumption vs. block size; (c) transaction confirmation rate vs. number of miners; (d) transaction confirmation rate vs. block size; (e) miners' average reward vs. number of miners; and (f) miners' average reward vs. block interval

ACKNOWLEDGMENT

This paper is partly supported by Academy of Finland via: (a) FIREMAN consortium CHIST-ERA-17-BDSI-003/n.326270, (b) EnergyNet Research Fellowship n.321265/n.328869 and (c) 6G Flagship n. 346208; and by Jane and Aatos Erkko Foundation via STREAM project. We would like to thank Prof. Julian Cheng from University of British Columbia for his valuable comments during the paper preparation, Dr. Hanna Niemelä for proofreading this paper, and Arthur Sousa de Sena for producing the graphics of Figs. 1 and 2.ACKNOWLEDGMENT

This paper is partly supported by Academy of Finland via: (a) FIREMAN consortium CHIST-ERA-17-BDSI-003/n.326270, (b) EnergyNet Research Fellowship n.321265/n.328869 and (c) 6G Flagship n. 346208; and by Jane and Aatos Erkko Foundation via STREAM project. We would like to thank Prof. Julian Cheng from University of British Columbia for his valuable comments during the paper preparation, Dr. Hanna Niemelä for proofreading this paper, and Arthur Sousa de Sena for producing the graphics of Figs. 1 and 2.

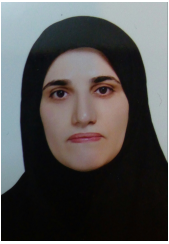

**Shiva Kazemi Taskou** is a PhD candidate at the Department of Computer Engineering, Amirkabir University of Technology, Tehran, Iran. Her current research interests include resource management in wireless networks, wireless network virtualization, machine learning, deep learning, and optimization.

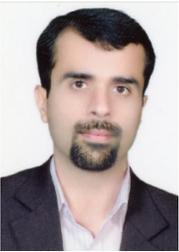

**Mehdi Rasti** is currently an Associated Professor at the Department of Computer Engineering, Amirkabir University of Technology, Tehran, Iran and is a visiting researcher at the Lappeenranta-Lahti University of Technology (LUT), Lappeenranta, Finland. His current research interests include radio resource allocation in IoT, Beyond 5G and 6G wireless networks.

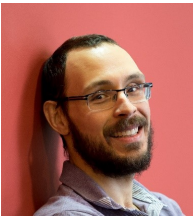

**Pedro H. J. Nardelli** is an Associate Professor (tenure track) in IoT in Energy Systems at LUT University, Finland. He is also Academy of Finland Research Fellow and Strategic Vertical Area coordinator for Energy in the 6G Flagship at University of Oulu, Finland.